\documentclass[12pt]{article}
\usepackage{graphicx}
\begin{document}

\title{\bf Intrasubband plasmons in a finite array of quantum wires placed into an external magnetic field}

\author{Bludov Y.V.}
\date{}
\maketitle

\begin{center}
{\it \small Usikov Institute for Radiophysics and
Electronics, NAS of Ukraine\\ Acad. Proscura st., 12, Kharkov,
61085, Ukraine\\
Email: bludov@ire.kharkov.ua}
\end{center}

\begin{center} \textbf{Abstract} \end{center}
The paper deals with the theoretical investigation of intrasubband plasmons in an array of quantum wires, consisting of
a finite number of quantum wires, arranged at an equal distance from each other and placed into an external magnetic
field. Two types of quantum wire array are under consideration: an ordered array of quantum wires with equal electron
densities in all quantum wires and weakly disordered array of quantum wires, which is characterized by the fact that
the density of electrons of one defect quantum wire was different from that of other quantum wires. For the ordered
array of quantum wires, placed into the external magnetic field, the nonmonotonical dependence of plasmon frequency
upon the 1D density of electrons in quantum wires is predicted. For high magnetic field the existence of 1D electron
density ranges, in which plasmon modes do not exist, is shown. For the weakly disordered array of quantum wires the
existence of the local plasmon modes, which properties differ from those of usual modes, is found. At high magnetic
field the disappearance of the local plasmon modes at certain ranges of 1D electron density in defect quantum wire is
shown.

\section{Introduction}

Quasi one-dimensional electron systems (1DES) or quantum wires (QW) are artificial struc\-tu\-res in which the motion
of charge carriers is confined in two transverse directions but is essentially free (in the effective mass sense) in
the longitudinal direction \cite{kn1,kn2,kn3}. Usually QW are produced by adding an additional one-dimensional
confinement of a two-dimensional electron system (2DES). This additional confinement is, in general, weaker than the
strong confinement of original 2DES \cite{rebpro}. One of motivations to study QW is the fact that the mobility of
charge carriers is higher than in 2DES on which they are built. The reason for this is that the impurity content and
distribution around QW can be selectively controlled, producing enhanced mobility \cite{dsfirst}.

Collective charge-density excitations, or plasmons in quantum wires (QW) are of great interest to physicists. Earlier
plasmons in QW were investigated both theoretically \cite{dsfirst,dsb,gold,ldj,hw} and experimentally
\cite{han,dem,dem1,goni}. In those papers it was shown that plasmons in QW possess some new unusual dispersion
properties. Firstly, the plasmon spectrum strongly depends on the width of QW. Secondly, 1D plasmons are free from the
Landau damping \cite{dsb,hw} in the whole range of wavevectors.

From the point of view of practical application the so-called
weakly disordered arrays of low-dimensional systems are the
objects of interest. Recently the plasmons in weakly disordered
superlattice, formed of a finite number of equally spaced
two-dimensional electron systems, have been theoretically
investigated in cases where the external magnetic field is absent
\cite{gvozd,js,sy} or present \cite{ya}. The weakly disordered
superlattice is characterized by the fact that all of
two-dimensional systems possess the equal density of electrons
except one ("defect") two-dimensional system, whose density of
electrons differs from that of other two-dimensional systems. It
was found that the plasmon spectrum of such an array contains the
local plasmon mode, whose properties differ from those of other
plasmon modes. The existence of local plasmon mode is completely
analogous to the existence of local phonon mode, first obtained by
Lifshitz in 1947 for the problem of the phonon modes in a regular
crystal containing a single isotope impurity \cite{lif}. Notice
that practically all flux of electromagnetic energy of plasmons,
which correspond to the local mode, are concentrated in the
vicinity of defect 2DES. At the same time the paper \cite{ya}
indicated the opportunity of using the plasmon spectrum
peculiarities to determine the parameters of defects in the
superlattice.

Plasmons in a finite weakly disordered array of QWs without
external magnetic field have been investigated theoretically in
paper \cite{fnt}. It has been supposed that the defect QW can
occupy an arbitrary position in the array. It was shown in paper
\cite{fnt}
that the position of the defect
QW in the array does not affect strongly the spectrum of the local
plasmon mode but it exerts an significant influence on the
spectrum of other plasmon modes. At the same time, when the defect
QW is arranged inside the array, the plasmon spectrum contains
modes, whose dispersion properties do not depend on the value of
the electron density in the defect QW.

The external magnetic field is known to cause considerable changes
in plasmon spectrum of low-dimensional structures. So, earlier
plasmons in single two-dimensional electron system placed into
external magnetic field directed perpendicularly to 2DES, was
investigated both theoretically \cite{74} and experimentally
\cite{75}. It was shown, that the dispersion relation for plasmons
in 2DES placed into the external magnetic field can be expressed
as
\begin{equation}
\omega^2_H=\omega_c^2+\omega^2, \label{pwmf}
\end{equation}
where $\omega_H$ is the frequency of plasmon in presence of
external magnetic field, $\omega_c=eB/m^*c$ is the frequency of
cyclotron resonance, $\omega$ is the frequency of plasmons in the
case where external magnetic field is absent.

Plasmons in single 1DES was also investigated theoretically \cite{21,kit} and experimentally \cite{dem,dem1}. As it was
shown experimentally \cite{dem}, the dispersion law for one-dimensional plasmon in the presence of magnetic field also
can be described by the dependence (\ref{pwmf}). Nevertheless in paper \cite{dem1} another one-dimensional plasmon mode
was found experimentally. The later mode possesses the negative magnetic field dispersion. At the same time paper
\cite{21} shows theoretically, that the above-mentioned negative magnetic field dispersion in the case of one
dimensionality occurs in high magnetic field only. At weak magnetic field the properties of intrasubband plasmons in
single QW depend considerably upon the value of one-dimensional electrons density in QW. Thus, if density of electrons
in QW exceeds a certain critical value, the intrasubband plasmon frequency increases as the magnetic field increases.
In the opposite case, when density of electrons in QW is smaller then the critical value, the intrasubband plasmon
frequency decreases as the magnetic field increases.

In this paper we investigate intrasubband plasmons in a finite array of QWs, placed into an external magnetic field. We
consider two types of QW array: an array in which the 1D elecron densities are equal in all QWs (ordered array of QWs)
and an array in which the 1D electron density of one defect QW differs from that of other QWs (weakly disordered array
of QWs). The paper is organized as follows. In section 2 we derive the dispersion relation for plasmons in the finite
array of QWs. In section 3 we present the results of the numerical solution of the dispersion relation for intrasubband
plasmons in ordered array of QWs. In section 4 we discuss the dispersion properties of intrasubband plasmons in the
weakly disordered array of QWs. We conclude the paper with a brief summary of results and possible applications
(section 5).

\section{Dispersion relation}

We consider the array of QWs consisting of a finite number $M$ of QWs, arranged at planes $z=ld$ ($l=0,...,M-1$ is the
number of QW, $d$ is the distance between adjacent QWs). At the same time we suppose that the 1D density of electrons
in $l$-th QW is equal to $N_l$. QW are considered to be placed into the uniform dielectric medium with the dielectric
constant $\varepsilon$. We use such a simple model (in which the dielectric constants of the media inside and outside
the array are equal) to avoid the appearance of a surface plasmon mode. We consider the movement of electrons to be
free in the $x$-direction and is considerably confined in the directions $y$ and $z$. We assume that the array of QWs
is built on an ideal 2DES by application an additional confining potential along $y$-direction, which are parabolic:
$U_{conf}=\frac{1}{2}m^*\omega_0^2y^2$. Here $m^*$ is the effective mass of electron, $\omega_0$ is the classical
oscillation frequency of electron, placed in potential $U_{conf}$. At the same time we suppose that the width of all
QWs is equal to zero in $z$-direction. The external constant magnetic field is considered to be directed
perpendicularly to plane $xy$ along axis $z$.

To obtain the single-particle wave-function of the electron in QW
we write down the expression for vector potential $\textbf{A}$ in
Landau gauge: $\textbf{A}=(-By,0,0)$. So, in that case the
single-particle Hamiltonian of the electrons can be represented as
follows
\begin{equation}
\hat{H}=\frac{1}{2m^*}\left(\hat{\textbf{p}}+\frac{e}{c}\textbf{A}\right)^2+U_{conf}(x,y),
\label{h0}
\end{equation}
where $\hat{\textbf{p}}=-i\hbar\nabla$ is the operator of the
momentum of electron. In expression (\ref{h0}) we neglect the spin
splitting in the magnetic field.

We seek an explicit form for the electron wave-function:
$\psi(x,y)=\exp(ikx)\phi(y)$. In this case the
Schr\"{o}dinger equation $\hat{H}\psi(x,y)=E\psi(x,y)$ after
some algebra can be written as
\begin{equation}
-\frac{\hbar^2}{2m^*}\frac{\partial^2\phi}{\partial
y^2}+\frac{1}{2}m^*\Omega^2\left[y-\alpha
k\right]^2\phi(y)=\left\{E-\frac{\hbar^2k^2}{2m^*}\frac{\omega_0^2}{\Omega^2}\right\}\phi(y),
\label{sch-eq}
\end{equation}
where $\alpha=\hbar\omega_c/m^*\Omega^2$,
$\Omega^2=\omega_c^2+\omega_0^2$. The solution of equation
(\ref{sch-eq}) is a shifted harmonic oscillator wave
function. So, the expression for energy subbands and a
single-particle wave function for the electron in $l$-th QW
can be written in the form \cite{6}:
\begin{equation}
E_m(k)=E_m+(1/2m^*)(\omega_0/\Omega)^2\hbar^2k^2, \label{en-sb}
\end{equation}
\begin{equation}
\psi_{l,m,k}({\bf r})=(1/2\pi)^{1/2}e^{ikx}\phi_m(y-\alpha
k)\left[\delta(z-ld)\right]^{1/2}. \label{wf}
\end{equation}
Here
\begin{equation}
E_m=\hbar\Omega(m+1/2), \qquad
\phi_m(y)=(2^mm!\pi^{1/2}l_{\Omega})^{-1/2}\exp\left(-\frac{y^2}{2l_{\Omega}^2}\right)H_m(y/l_{\Omega}),
\end{equation}
$m$ is the number of energy subband, $H_m(y)$ is an Hermite
polynomial, $l_{\Omega}=(\hbar/m^*\Omega)^{1/2}$ is a typical
width of wave function (which is simply a magnetic length, if
$\omega_0=0$).

As it can be seen from the expression (\ref{en-sb}), in the
presence of confining potential in $y$-direction the degeneracy of
Landau-levels is broken and each Landau-level forms subband. At
the same time the wave function (\ref{wf}) in $y$-direction
depends on the wavevector $k$ in $x$-direction. So, in the
presence of confining potential and external magnetic fields the
directions $x$ and $y$ are coupled.

To obtain the collective excitations spectrum we start with a standard linear-response theory in an random phase
approximation. Let us consider $\delta n(\bf{r})$ which is the deviation of the electron density from its equilibrium
value. After using the standard linear-response theory and the random phase approximation, the matrix element of the
electron density deviation from its equilibrium value $\delta n_{\alpha,\alpha'}=\langle\alpha|\delta
n|\alpha'\rangle=\int d{\bf r}\psi_{\alpha}^*({\bf r})\psi_{\alpha^{\prime}}({\bf r})\delta n(\bf{r})$ can be related
to the perturbation as
\begin{equation}
\delta
n_{\alpha\alpha'}=\frac{f_{\alpha'}-f_{\alpha}}{E_{\alpha'}-E_{\alpha}+\hbar\omega}V_{\alpha\alpha'}.
\label{eq:nv}
\end{equation}
Here $\alpha=(l,m,k)$ is a composite index, $f_{\alpha}$ is the
Fermi distribution function,
$V_{\alpha,\alpha'}=\langle\alpha|V|\alpha'\rangle$ are the matrix
elements of the perturbing potential $V=V^{ex}+V^H$, $V^{ex}$ and
$V^H$ are the external and Hartree potentials, respectively.

Note, that the matrix elements of Hartree potential can be expressed through the per\-tur\-bation \cite{dsb} as
\begin{equation}
\displaystyle{V^H_{\alpha \alpha^{\prime}}=\frac{e^2}%
{\varepsilon}\int d{\bf r}\psi_{\alpha}^*({\bf
r})\psi_{\alpha^{\prime}}({\bf r})\int \frac{d{\bf r_1}}{|{\bf
r}-{\bf r}_1|}\delta n({\bf r}_1).} \label{me}
\end{equation}
Taking into account, that $$ \delta n({\bf r}_1)=\sum_{\beta
\beta^{\prime}}\delta
n_{\beta\beta^{\prime}}\psi_{\beta^{\prime}}^*({\bf
r}_1)\psi_{\beta}({\bf r}_1),\quad \beta=(n,s,k_1)$$ we obtain
\begin{equation}
V^H_{\alpha \alpha^{\prime}}=\sum_{\beta \beta^{\prime}}W_{\alpha
\alpha^{\prime}\beta\beta^{\prime}}\delta n_{\beta\beta^{\prime}}.
\label{eq:vw}
\end{equation}
Here
\begin{eqnarray}
W_{\alpha\alpha^{\prime}\beta\beta^{\prime}}&=&\frac{e^2}%
{\varepsilon}\int d{\bf r}\psi_{\alpha}^*({\bf r})\psi_{\alpha^{\prime}}({\bf r})\int \frac{d{\bf r_1}}{|{\bf r}-{\bf
r}_1|}\psi_{\beta^{\prime}}^*({\bf r}_1)\psi_{\beta}({\bf r}_1)= \nonumber
\\ &=&\frac{\delta(q-q_1)}{2\pi}%
\frac{2e^2}{\varepsilon}W_{l,n|m,m',s,s'}\left(k^{\prime},k,k_1^{\prime},k_1\right)\delta_{n,n^{\prime}}%
\delta_{l,l^{\prime}}, \label{eq:wall}
\end{eqnarray}
where
\begin{eqnarray}W_{l,n|m,m',s,s'}\left(k^{\prime},k,k_1^{\prime},k_1\right)=&&\int \phi_m(y-\alpha k)\phi_{m'}(y-\alpha
k^{\prime})\phi_{s'}(y_1-\alpha k_1^{\prime})\phi_s(y_1-\alpha k_1)\times\nonumber\\ &&\quad\times
K_0\left(q((y-y_1)^2+(l-n)^2d^2)^{1/2}\right)dydy_1 ,\label{eq:www}\end{eqnarray} $q=k^{\prime}-k$,
$q_1=k_1^{\prime}-k_1$, $K_0(x)$ is the zeroth-order modified Bessel function of the second kind. From the equations
(\ref{eq:nv}),(\ref{eq:vw}),(\ref{eq:wall}),after some algebra we get
\begin{eqnarray}
\delta n_{lmk,lm^{\prime}k+q}&=&\frac{f_{lm^{\prime}k+q}-f_{lmk}}%
{E_{lm^{\prime}k^{\prime}}-E_{lmk}+\hbar\omega}\left(V^{ex}_{lmk,lm^{\prime}k^{\prime}}+\right.\nonumber\\%
&&+\left.\frac{2e^2}{\varepsilon}\frac{1}{\pi}\sum_{n,s,s^{\prime}}%
\int dk_1
W_{l,n|m,m^{\prime},s,s^{\prime}}\left(k+q,k,k_1+q,k_1\right)\delta
n_{nsk_1,ns^{\prime}k_1+q}\right). \label{eq:nras}
\end{eqnarray}
The factor of 2 before summation comes from the spin degeneracy.

Collective excitations of QW array exist when equation (\ref{eq:nras}) has a nonzero solution $\delta n$ in the case
where the external perturbation $V^{ex}=0$. Since the parameter $\alpha k$ is the small value \cite{21}, we can expand
the wave function in terms of $\alpha$ as $\phi_m(y-\alpha k)=\phi_m(y)-\alpha
k\phi^{\prime}_m(y)+\frac{1}{2}\alpha^2k^2\phi^{\prime\prime}_m(y)$. Besides that, taking into account, that when $q
\to 0$, we can admit $\alpha(k+q) \approx \alpha k$, $\alpha(k_1+q) \approx \alpha k_1$. Under this assumption we can
represent expression (\ref{eq:www}) in the form \begin{eqnarray}
W_{l,n|m,m^{\prime},s,s^{\prime}}\left(k+q,k,k_1+q,k_1\right)&\approx&
C_{l,n|m,m^{\prime},s,s^{\prime}}^{(0)}(q)+\alpha kC_{l,n|m,m^{\prime},s,s^{\prime}}^{(1)}(q)+\nonumber\\
&&+\alpha^2k^2C_{l,n|m,m^{\prime},s,s^{\prime}}^{(2)}(q)+k_1\left\{\alpha
B_{l,n|m,m^{\prime},s,s^{\prime}}^{(1)}(q)+\right.\label{eq:det-wln}\\
&&\left.+\alpha^2kB_{l,n|m,m^{\prime},s,s^{\prime}}^{(2)}(q)\right\}+%
\alpha^2k_1^2A_{l,n|m,m^{\prime},s,s^{\prime}}(q),\nonumber
\end{eqnarray}
where \begin{eqnarray} C_{l,n|m,m^{\prime},s,s^{\prime}}^{(0)}(q)&=&\int
\phi_m(y)\phi_{m^{\prime}}(y)\phi_s(y_1)\phi_{s^{\prime}}(y_1)K_0\left(q((y-y_1)^2+(l-n)^2d^2)^{1/2}\right)dydy_1,\nonumber\\
C_{l,n|m,m^{\prime},s,s^{\prime}}^{(1)}(q)&=&-\int
\left\{\phi^{\prime}_m(y)\phi_{m^{\prime}}(y)+\phi^{\prime}_{m^{\prime}}(y)\phi_m(y)\right\}\times\nonumber\\
&&\quad\times\phi_s(y_1)\phi_{s^{\prime}}(y_1)K_0\left(q((y-y_1)^2+(l-n)^2d^2)^{1/2}\right)dydy_1,\nonumber\\
C_{l,n|m,m^{\prime},s,s^{\prime}}^{(2)}(q)&=&\frac{1}{2}\int
\left\{\phi^{\prime\prime}_m(y)\phi_{m^{\prime}}(y)+\phi^{\prime\prime}_{m^{\prime}}(y)\phi_m(y)+%
2\phi^{\prime}_m(y)\phi^{\prime}_{m^{\prime}}(y)\right\}\times\nonumber\\
&&\quad\times\phi_s(y_1)\phi_{s^{\prime}}(y_1)K_0\left(q((y-y_1)^2+(l-n)^2d^2)^{1/2}\right)dydy_1,\nonumber\\
B_{l,n|m,m^{\prime},s,s^{\prime}}^{(1)}(q)&=&-\int
\left\{\phi^{\prime}_s(y_1)\phi_{s^{\prime}}(y_1)+\phi^{\prime}_{s^{\prime}}(y_1)\phi_s(y_1)\right\}\times\nonumber\\
&&\quad\times\phi_m(y)\phi_{m^{\prime}}(y)K_0\left(q((y-y_1)^2+(l-n)^2d^2)^{1/2}\right)dydy_1,\nonumber\\
B_{l,n|m,m^{\prime},s,s^{\prime}}^{(2)}(q)&=&\int
\left\{\phi^{\prime}_m(y)\phi_{m^{\prime}}(y)+\phi^{\prime}_{m^{\prime}}(y)\phi_m(y)\right\}\times\nonumber\\
&&\quad\times\left\{\phi^{\prime}_s(y_1)\phi_{s^{\prime}}(y_1)+\phi^{\prime}_{s^{\prime}}(y_1)\phi_s(y_1)\right\}K_0\left(q((y-y_1)^2+(l-n)^2d^2)^{1/2}\right)dydy_1,\nonumber\\
A_{l,n|m,m^{\prime},s,s^{\prime}}(q)&=&\frac{1}{2}\int
\left\{\phi^{\prime\prime}_s(y_1)\phi_{s^{\prime}}(y_1)+\phi^{\prime\prime}_{s^{\prime}}(y_1)\phi_s(y_1)+%
2\phi^{\prime}_s(y_1)\phi^{\prime}_{s^{\prime}}(y_1)\right\}\times\nonumber\\
&&\quad\times\phi_m(y)\phi_{m^{\prime}}(y)K_0\left(q((y-y_1)^2+(l-n)^2d^2)^{1/2}\right)dydy_1.\nonumber
\end{eqnarray}
Substituting the expression (\ref{eq:det-wln}) into
(\ref{eq:nras}), we obtain: \begin{eqnarray}
\delta n_{lmk,lm^{\prime}k+q}=&&\frac{2e^2}{\varepsilon}\frac{1}{\pi}\frac{f_{lm^{\prime}k+q}-f_{lmk}}%
{E_{lm^{\prime}k+q}-E_{lmk}+\hbar\omega}\times\label{eq:nras3}\\ &&\times\sum_{n,s,s^{\prime}}\int dk_1 \delta
n_{nsk_1,ns^{\prime}k_1+q}\left[C_{l,n|m,m^{\prime},s,s^{\prime}}^{(0)}(q)+\alpha
kC_{l,n|m,m^{\prime},s,s^{\prime}}^{(1)}(q)+\right.\nonumber\\
&&+\alpha^2k^2C_{l,n|m,m^{\prime},s,s^{\prime}}^{(2)}(q)+k_1\left\{\alpha
B_{l,n|m,m^{\prime},s,s^{\prime}}^{(1)}(q)+\alpha^2kB_{l,n|m,m^{\prime},s,s^{\prime}}^{(2)}(q)\right\}+\nonumber\\
&&\left.+\alpha^2k_1^2A_{l,n|m,m^{\prime},s,s^{\prime}}(q)\right]\nonumber.
\end{eqnarray}
After multiplication both left and right hand side of
equation (\ref{eq:nras3}) by $2k^i$ ($i=0,1,2$) and
integration, we get:
\begin{eqnarray}
\chi_{l|m,m^{\prime}}^{(0)}=\frac{2e^2}{\varepsilon}\sum_{n,s,s^{\prime}}&&%
\left[\left\{C_{l,n|m,m^{\prime},s,s^{\prime}}^{(0)}(q)\Pi_{l|m,m^{\prime}}^{(0)}+\alpha
C_{l,n|m,m^{\prime},s,s^{\prime}}^{(1)}(q)\Pi_{l|m,m^{\prime}}^{(1)}+\right.\right.\label{eq:osur1}\\
&&\left.\left.+\alpha^2C_{l,n|m,m^{\prime},s,s^{\prime}}^{(2)}(q)\Pi_{l|m,m^{\prime}}^{(2)}\right\}\chi_{n|s,s^{\prime}}^{(0)}+\right.\nonumber\\
&&+\left.\left\{\alpha
B_{l,n|m,m^{\prime},s,s^{\prime}}^{(1)}(q)\Pi_{l|m,m^{\prime}}^{(0)}+\alpha^2B_{l,n|m,m^{\prime},s,s^{\prime}}^{(2)}(q)\Pi_{l|m,m^{\prime}}^{(1)}\right\}\chi_{n|s,s^{\prime}}^{(1)}+\right.\nonumber\\
&&+\left.\alpha^2A_{l,n|m,m^{\prime},s,s^{\prime}}(q)\Pi_{l|m,m^{\prime}}^{(0)}\chi_{n|s,s^{\prime}}^{(2)}\right],\nonumber\\
\chi_{l|m,m^{\prime}}^{(1)}=\frac{2e^2}{\varepsilon}\sum_{n,s,s^{\prime}}&&%
\left[\left\{C_{l,n|m,m^{\prime},s,s^{\prime}}^{(0)}(q)\Pi_{l|m,m^{\prime}}^{(1)}+\alpha
C_{l,n|m,m^{\prime},s,s^{\prime}}^{(1)}(q)\Pi_{l|m,m^{\prime}}^{(2)}+\right.\right.\\
&&\left.\left.+\alpha^2C_{l,n|m,m^{\prime},s,s^{\prime}}^{(2)}(q)\Pi_{l|m,m^{\prime}}^{(3)}\right\}\chi_{n|s,s^{\prime}}^{(0)}+\right.\nonumber\\
&&+\left.\left\{\alpha
B_{l,n|m,m^{\prime},s,s^{\prime}}^{(1)}(q)\Pi_{l|m,m^{\prime}}^{(1)}+\alpha^2B_{l,n|m,m^{\prime},s,s^{\prime}}^{(2)}(q)\Pi_{l|m,m^{\prime}}^{(2)}\right\}\chi_{n|s,s^{\prime}}^{(1)}+\right.\nonumber\\
&&+\left.\alpha^2A_{l,n|m,m^{\prime},s,s^{\prime}}(q)\Pi_{l|m,m^{\prime}}^{(1)}\chi_{n|s,s^{\prime}}^{(2)}\right],\nonumber\\
\chi_{l|m,m^{\prime}}^{(2)}=\frac{2e^2}{\varepsilon}\sum_{n,s,s^{\prime}}&&%
\left[\left\{C_{l,n|m,m^{\prime},s,s^{\prime}}^{(0)}(q)\Pi_{l|m,m^{\prime}}^{(2)}+\alpha
C_{l,n|m,m^{\prime},s,s^{\prime}}^{(1)}(q)\Pi_{l|m,m^{\prime}}^{(3)}+\right.\right.\label{eq:osur3}\\
&&\left.\left.+\alpha^2C_{l,n|m,m^{\prime},s,s^{\prime}}^{(2)}(q)\Pi_{l|m,m^{\prime}}^{(4)}\right\}\chi_{n|s,s^{\prime}}^{(0)}+\right.\nonumber\\
&&+\left.\left\{\alpha
B_{l,n|m,m^{\prime},s,s^{\prime}}^{(1)}(q)\Pi_{l|m,m^{\prime}}^{(2)}+\alpha^2B_{l,n|m,m^{\prime},s,s^{\prime}}^{(2)}(q)\Pi_{l|m,m^{\prime}}^{(3)}\right\}\chi_{n|s,s^{\prime}}^{(1)}+\right.\nonumber\\
&&+\left.\alpha^2A_{l,n|m,m^{\prime},s,s^{\prime}}(q)\Pi_{l|m,m^{\prime}}^{(2)}\chi_{n|s,s^{\prime}}^{(2)}\right]\nonumber,
\end{eqnarray}
where $$\chi_{n|s,s^{\prime}}^{(i)}=2\int dk_1 k_1^i\delta
n_{nsk_1,ns^{\prime}k_1+q}, \quad
\chi_{l|m,m^{\prime}}^{(i)}=2\int dk k^i\delta
n_{lmk,lm^{\prime}k+q},$$ $$\Pi_{l|m,m^{\prime}}^{(i)}=\frac{1}{\pi}\int dk\,k^i\frac{f_{lm^{\prime}k+q}-f_{lmk}}%
{E_{lm^{\prime}k+q}-E_{lmk}+\hbar\omega}.$$

We restrict our consideration to the case of intrasubband plasmons. So, we consider that the intersubband trasitions of
charge carriers are absent. In this case $\Pi_{l|m,m^{\prime}}^{(i)}=0$, if $m \ne m^{\prime}$ and the system of
equations (\ref{eq:osur1})--(\ref{eq:osur3}) can be rewritten as follows
\begin{equation}
\chi_{l|m,m}^{(p)}=\frac{2e^2}{\varepsilon}\sum_{n,s,t}U_{l,n,m,s,p,t}\chi_{n|s,s}^{(t)},
\label{eq:ocom}
\end{equation}
where $$U_{l,n,m,s,p,1}=C_{l,n|m,m,s,s}^{(0)}(q)\Pi_{l|m,m}^{(p)}+\alpha
C_{l,n|m,m,s,s}^{(1)}(q)\Pi_{l|m,m}^{(p+1)}+\alpha^2C_{l,n|m,m,s,s}^{(2)}(q)\Pi_{l|m,m}^{(p+2)},$$
$$U_{l,n,m,s,p,2}=\alpha B_{l,n|m,m,s,s}^{(1)}(q)\Pi_{l|m,m}^{(p)}+\alpha^2B_{l,n|m,m,s,s}^{(2)}(q)\Pi_{l|m,m}^{(p+1)},
$$ $$U_{l,n,m,s,p,3}=\alpha^2A_{l,n|m,m,s,s}(q)\Pi_{l|m,m}^{(p)}, \quad p=0,...,2.$$ Equation (\ref{eq:ocom}) is a set
of linear equation and it has nonearo solution in the case where its determinant is equal to zero. So, the plasmon
dispersion relation can be written in the form:
\begin{equation}
{\rm
det}\left\|\delta_{l,n}\delta_{m,s}\delta_{p,t}-\frac{2e^2}{\varepsilon}U_{l,n,m,s,p,t}\right\|=0.
\label{eq:dre}
\end{equation}
Note, that as $M=1$, the dispersion relation (\ref{eq:dre}) coincides with the dispersion relation for plasmons in
single QW in the presence of external magnetic field obtained in \cite{21}.

At a zero temperature and in the long-wavelength limit (where $q \to 0$) function $\Pi_{l|m,m}^{(i)}$ can be written as
\begin{eqnarray}
&\Pi_{l|m,m}^{(0)}=\displaystyle{\frac{2g_m^l}{\pi
m_r}\frac{q^2}{\omega^2},\quad
\Pi_{l|m,m}^{(1)}=-\frac{a}{b}\Pi_{l|mm}^{(0)},}&\nonumber\\
&\displaystyle{\Pi_{l|m,m}^{(2)}=\left(\frac{a}{b}\right)^2\Pi_{l|mm}^{(0)}-\frac{q}{b}\frac{2g_m^l}{\pi},}\quad
\displaystyle{\Pi_{l|m,m}^{(3)}=-\left(\frac{a}{b}\right)^3\Pi_{l|mm}^{(0)}+\frac{q}{b}\frac{2g_m^l}{\pi}\left(q+\frac{a}{b}\right),}&\nonumber\\
&\displaystyle{\Pi_{l|m,m}^{(4)}=\left(\frac{a}{b}\right)^4\Pi_{l|mm}^{(0)}-\frac{q}{b}\frac{2g_m^l}{\pi}\left[q^2+\left(\frac{2g_m^l}{\pi}\right)^2+%
\frac{a}{b}q+\left(\frac{a}{b}\right)^2\right],}&\nonumber
\end{eqnarray}
where $$a=\frac{\hbar^2q^2}{2m_r}+\hbar\omega,\qquad b=\frac{\hbar^2q}{m_r}, \qquad m_r=m^*(\Omega/\omega_0)^2, \qquad
g^l_m=\frac{1}{\hbar}\sqrt{2m_r(E_F^{l}-E_m)},$$ $E_F^{l}$ is the Fermi level in $l$-th QW.

\section{Intrasubband plasmons in the ordered QW array}

Fig.1 presents the dispersion curves for intrasubband plasmons in finite ordered array of QWs (in which 1D electron
densities are equal in all QWs), placed into the external magnetic field. The $y$-axis gives the dimensionless
frequency $\omega/\omega_0$, and the $x$-axis gives the dimensionless wavevector $ql_0$. As the model of QW we use
heterostructure GaAs with the effective mass of electrons $m^*=0,067m_0$ ($m_0$ is the mass of free electron) and the
dielectric constant $\varepsilon=12$. For comparison the dispersion curve for the plasmons in a single QW with the same
parameters is depicted in Fig.1 by dashed curve 1. As seen from Fig.1, the intrasubband plasmon spectrum in the finite
ordered array of QWs contains $M$ modes. Thus, the number of modes in the spectrum is equal to the number of QWs in the
array \cite{gvozd} (it should be mentioned that in the case under consideration the value of plasma frequency of the
electrons in QWs is chosen so, that in each QW only the lowest energy subband is occupied by electrons). Notice that
with an increase of wavenumber $q$ the plasmon frequency $\omega$ increases monotonically likewise. It should be
emphasized that in the limit $qd \to \infty$, when the Coulomb interaction between electrons in adjacent QWs is
neglible, the dispersion curves for plasmon modes are gradually drawn together and are close to the dispersion curve
for the plasmon in the single QW with the same density of electrons (dashed curve 1).

Now we consider the influence of the electron density value on the properties of in\-tra\-sub\-band plasmons in ordered
array of QWs. Fig.2 presents the dependence of plasmon frequency upon the plasma frequency of electrons in QWs in the
case of the fixed value of wavevector $q$ and for different values of cyclotron frequency of electrons. The $y$-axis
gives the dimensionless frequency $\omega/\omega_0$, and the $x$-axis gives the dimensionless plasma frequency of
electrons in QWs $\omega_p/\omega_0$. We consider first the case when $\omega_c=0$ (Fig.2a), i.e. when the external
magnetic field is absent. As Fig.2a shows, in this case the frequency of intrasubband plasmons increases when the value
of plasma frequency of electrons in QWs $\omega_p$ is increased. At the same time at small values of $\omega_p$, when
the Fermi energy is below the bottom of first subband ($E_F^l<E_1$) and only the lowest (zero) subband in each QW is
occupied by electrons, the intrasubband plasmon spectrum contains $M$ modes (curves 1). Nevertheless when the value of
plasma frequency of electrons in QWs exceeds the value of $\omega_p \approx 2.68\omega_0$, the intrasubband plasmon
spectrum contains $2M$ modes (curves 1 and 2). In this case the Fermi energy is above the bottom of first subband but
below the bottom of second subband and consequently there are already two subbands (zero and first) in each QW, which
are occupied by electrons. With further increasing of $\omega_p$, new subbands become occupied by electrons, and each
occupied subband in each QW supports its own intrasunnabd plasmon. Hence the general number of intrasubband plasmon
modes in finite ordered array of QWs without an external magnetic field is equal to $nM$ ($n$ is the quantity of
subbands in each QW, occupied by electrons).

The properties of intrasubband plasmons somewhat change, when the ordered array of QWs is placed into the external
magnetic field. So, at weak magnetic field (Fig.2b), the frequency of intrasubband plasmons, supported by the lowest
subband (curves 1), is increased monotonically with the increasing of $\omega_p$. At the same time when the Fermi
energy exceeds the bottom of first subband (and it becomes populated by the electrons), the intrasubband plasmons,
supported by first subbands in each QW, arises in the spectrum (curves 2). The frequency of these plasmons (as distinct
from the case of zero magnetic field, see Fig.2a) increases nonmonotonically when the value of $\omega_p$ is increased.
So, starting with some value of $\omega_p$ (in our case starting with $\omega_p \approx 3.55\omega_0$) the frequency of
intrasubband plasmons, supported by first subbands, is decreased with the increasing of $\omega_p$ and when $\omega_p
\approx 4.0\omega_0$ these plasmons disappear. Notice that intrasubband plasmons, supported by second subbands (curves
3), possess the same properties.

In the case of higher magnetic field (Fig.2c) the dependence of intrasubband plasmon frequency upon the value of plasma
frequency of electrons in QWs possesses the following properties. So, in this case the frequency of intrasubband
plasmons, supported by zero (curves 1), first (curves 2) and second (curves 3) subbands depends nonmonotonically upon
the value of $\omega_p$. At the same time there are a certain intervals of values of $\omega_p$ (in this case
$2.7\omega_0<\omega_p<3.2\omega_0$, $4.45\omega_0<\omega_p<4.9\omega_0$), in which the intrasubband plasmons don't
exist.

\section{Intrasubband plasmons in a weakly disordered array of quantum wires}

We consider now the spectrum of intrasubband plasmons in weakly disordered array of QWs, in which all QWs possess the
equal 1D density of electrons $N$ except one defect QW whose density of electrons is equal to $N_{\rm{d}}$. So, the
density of electrons in $l$-th QW can be expressed as $N_l=(N_{\rm{d}}-N)\delta_{pl}+N$. Here $p$ is the number of
defect QW arranged at the plane $z=pd$, $\delta_{pl}$ is the Cronecker delta.

Fig.3 presents the spectrum of intrasubband plasmons (solid curves) in weakly disordered array of QWs for zero external
magnetic field. For comparison the dispersion curves for the intrasubband plasmons in a single QW with the electron
density $N$ and $N_{\rm{d}}$ are depicted by dashed curves 1 and 2, correspondingly. As seen from Fig.3, the
propagation of intrasubband plasmons in the weakly disordered array of QWs is characterized by the presence of the
local plasmon mode (LPM). At zero external magnetic field when the density of electrons in defect QW is less than the
density of electrons in other QW ($N_{\rm{d}}<N$), the LPM lies in the lower-frequency region in comparison with the
usual plasmon modes (Fig.3a). Accordingly, if $N_{\rm{d}}>N$, the LPM lies in the higher-frequency region in comparison
with the usual ones (Fig.3b) \cite{gvozd}. It should be emphasized that in the limit $qd \to \infty$, when the Coulomb
interaction between electrons in adjacent QWs is neglible, the LPM dispersion curve is close to the dispersion curve
for the plasmons in single QW with the density of electrons $N_{\rm{d}}$ (curve 2). Meanwhile, the dispersion curves
for usual plasmon modes in the limit $qd \to \infty$ are gradually drawn together and are close to the dispersion curve
for the plasmon in the single QW with the density of electrons $N$ (curve 1).

Now we consider the dependence of intrasubband plasmon spectrum upon the value of 1D electron density in the defect QW.
Fig.4 depicts the dependence of intrasubband plasmon frequency upon the ratio $N_{\rm{d}}/N$ for the fixed value of
wavevector $q$ and for different positions of the defect QW in the array. As seen from Fig.4, at zero external magnetic
field the number of LPM (depicted by bold solid curves) in the intrasubband plasmon spectrum is equal to the number of
subbands in defect QW, occupied by the electrons. As can be seen from the comparison of Fig.4a,b,c, the LPM spectrum is
weakly dependent upon the position of defect QW in the weakly disordered array of QWs. That phenomenon can be explained
by the fact, that practically the whole flux of the LPM electromagnetic energy is localized in the vicinity of the
defect QW \cite{fnt}. However, the spectrum of usual plasmon modes is more sensitive to the position of defect QW in
the array. Note that the frequency of LPM increases when the value of ratio $N_{\rm{d}}/N$ is increased.  At the same
time the usual plasmon modes spectrum is characterized by these features. As $p=0$ (Fig.4a) when the value of ratio
$N_{\rm{d}}/N$ is increased, the frequency of all usual plasmon modes increases as well. It should be noted that the
frequencies of intrasubband plasmons supported by first subbands of QWs (curves 1$'$--4$'$) are less sensitive to the
value of ratio $N_{\rm d}/N$ in comparison with the frequencies of intrasubband plasmons supported by zero subbands of
QWs (curves 1--4). However, when $p=1$ (Fig.4b) the frequencies of two of the usual plasmon modes (curves 2 and 2$'$)
does not practically depend upon the value of ratio $N_{\rm{d}}/N$. In the case where $p=2$ (Fig.4c) there are already
four intrasubband plasmon modes (curves 1,1$'$,3,3$'$) which possess such a distinctive feature. The spatial
distribution of the Hartree potential for those modes has a feature, that the absolute value of the Hartree potential
in the vicinity of the defect QW is neglible. Therefore, the defect QW does not exert a significant influence on the
dispersion properties of plasmon modes \cite{fnt}.

The properties of intrasubband plasmons change, if weakly disordered array of QWs is placed into the external magnetic
field. Fig.5 presents the dependence of plasmon frequency upon the value of ratio $N_d/N$ for fixed value of wavenumber
$q$ and for different positions of the defect QW in the array. As seen from Fig.5, at external magentic field the
dependence of LPM frequency upon ratio $N_d/N$ is nonmonotonic. So, the frequency of LPM, supported by defect QW zero
subband (curve LMP1), increases as the value of ratio $N_d/N$ is increased in the range $0<N_d/N<0.39$. At the same
time as the value of $N_d/N$ increases in the range $0.39<N_d/N<0.51$, the frequency of LMP, supported by defect QW
zero subband, is decreased. Meanwhile, the frequencies of LPM, supported by defect QW first and second subbands (curves
LPM2 and LPM3, correspondingly) also depend nonmonotonically upon the value of $N_d/N$. Notice that when weakly
disordered array of QWs is placed into an external magnetic field, there are certain ranges of 1D electron density in
defect QW (in Fig.5, e.g. $0.51<N_d/N<0.82$ and $1.46<N_d/N<1.98$), in which the LPMs don't exist. As seen from Fig.5
at external magnetic field (as in the case of zero external magnetic field) when $p=1$ and when $p=2$, the spectrum of
usual plasmon modes contains intrasubband modes (curve 2 in Fig.5b, curves 1 and 3 in Fig.5c), which frequencies don't
practically depend upon ratio $N_d/N$.

\section{Conclusion}

In conclusion, we calculated the intrasubband plasmon spectrum of the finite array of QWs placed into an external
magnetic field. Two types of QW arrays were under consideration: an ordered array of QWs (in which all the QWs possess
equal 1D density of electrons) and weakly disordered array of QWs (in which the 1D densities of electrons are equal in
all QWs except one defect QW). It is found that in the ordered array of QWs at zero magnetic field in each QW every
subbands, filled by electrons, support their own intrasubband plasmons. Hence the total quantity of intrasubband
plasmon modes in ordered QW array is equal to the number of QWs in the array multiplied by the number of filled
subbands in QW. Nevertheless, at nonzero external magnetic fields the quantity of intrasubband plasmon modes depends
upon the value of magnetic field and the 1D electron density of QWs. In particular, at high enough external magnetic
field there are certain ranges of 1D electron densities, in which none of intrasubband plasmon modes exists in the
spectrum.

In the case of weakly disordered array of QWs the LPMs whose properties differ from those of other modes exist in the
plasmon spectrum. We point out that as distinct from the case of zero magnetic field, at high enough magnetic field the
dependence of LPM frequency upon defect QW 1D density of electrons is of nonmonotonical character. Moreover, at high
magnetic field there are certain ranges of 1D electron density in defect QW, in which LMPs don't exist. At the same
time it is found that the intrasubband plasmon modes, whose spectrum does not depend upon the density of electrons of
the defect QW \cite{fnt}, exist also in the case of external magnetic field.

To conclude, it should be emphasized that the above-mentioned features of plasmon spectra can be used for the
diagnostics of defects in QW structures. Hence, the LPM properties can be used for the determination of the electron
density in the defect QW. At the same time the properties of usual plasmon modes can be used to define the position of
the defect QW in the array.

\pagebreak \

\pagebreak

\begin{center}
\textbf{Figures}
\end{center}

Figure 1. Dispersion curves of intrasubband plasmons in an ordered array of QWs with parameters: $M=5$, $d=15.0l_0$,
$N_l=N=const$ ($l=0,...,M-1$), $\omega_p=\left(2e^2N/\varepsilon m^*l_0^2\right)^{1/2}=1.5\omega_0$,
$\omega_c=0.75\omega_0$. \vspace{5mm}

Figure 2. The dependence of intrasubband plasmon frequency upon the plasma frequency of electrons in QW in the case
where $M=5$, $ql_0=0.04$, $d=15.0l_0$, $N_l=const$ ($l=0,...,M-1$) and for three values of cyclotron frequency of
electrons in QWs: $\omega_c=0$ (a), $\omega_c=0.25\omega_0$ (b), $\omega_c=0.75\omega_0$ (c). \vspace{5mm}

Figure 3. Disperison curves of intrasubband plasmons in weakly disordered array of QWs for parameters $M=5$,
$\omega_p=1.5\omega_0$, $\omega_c=0$, $d=15.0l_0$, $p=0$ and for two values of 1D density of electrons in defect QW:
$N_d/N=0,5$ (a), $N_d/N=1,5$ (b). The values of parameters are chosen in a manner that in all QWs only one (zeroth)
subband is occupied by electrons. \vspace{5mm}

Figure 4. The dependence of intrasubband plasmon frequency upon the ratio $N_d/N$ in the case where $M=5$, $ql_0=0.04$,
$d=15.0l_0$, $\omega_p=3.0\omega_0$, $\omega_c=0$ and for three different positions of defect QW in the array: $p=0$
(a), $p=1$ (b) and $p=2$ (c). At these values of parameters there are two subbands in all QWs (except the defect QW),
filled by electrons. Meanwhile, in defect QW the number of filled subbands is determined by the value of $N_{\rm d}$.
\vspace{5mm}

Figure 5. The dependence of intrasubband plasmon frequency upon the ratio $N_d/N$ in the case where $M=5$, $ql_0=0.04$,
$d=15.0l_0$, $\omega_p=3.2\omega_0$, $\omega_c=0.75$ and for different positions of defect QW in the array: $p=0$ (a),
$p=1$ (b) and $p=2$ (c). These values of parameters correspond to the fact that in all QWs (except the defect one) two
subbands (zeroth and first) are occupied by electrons. The number of filled subbands in defect QW is determined by the
value of $N_{\rm d}$.

\pagebreak

\begin{figure}[t!]
   \begin{center}
   \begin{tabular}{c}
   \includegraphics[270,216][515,391]{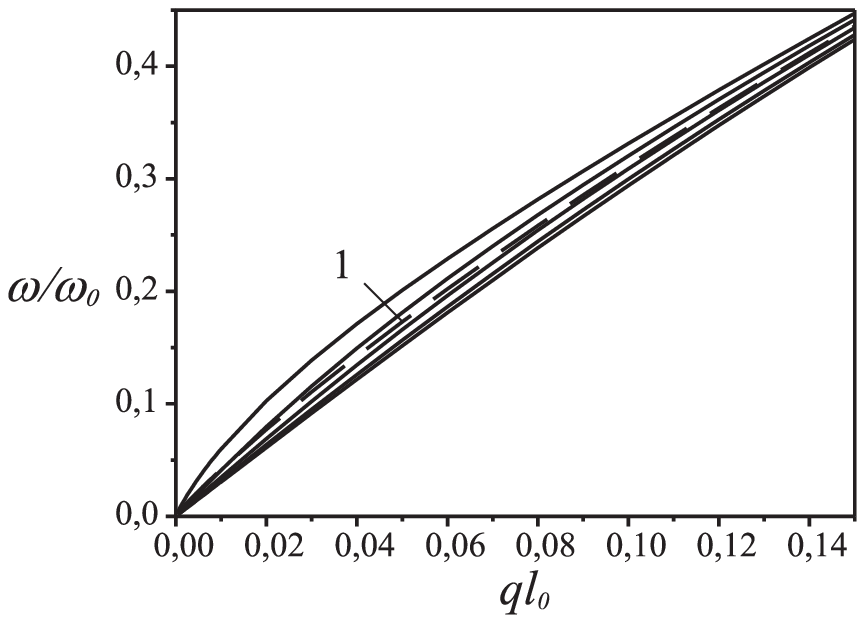}
   \end{tabular}
   \end{center}
\end{figure}
\begin{center} Figure 1. \end{center}

\pagebreak

\begin{figure}[t!]
   \begin{center}
   \begin{tabular}{c}
   \includegraphics[153,21][398,550]{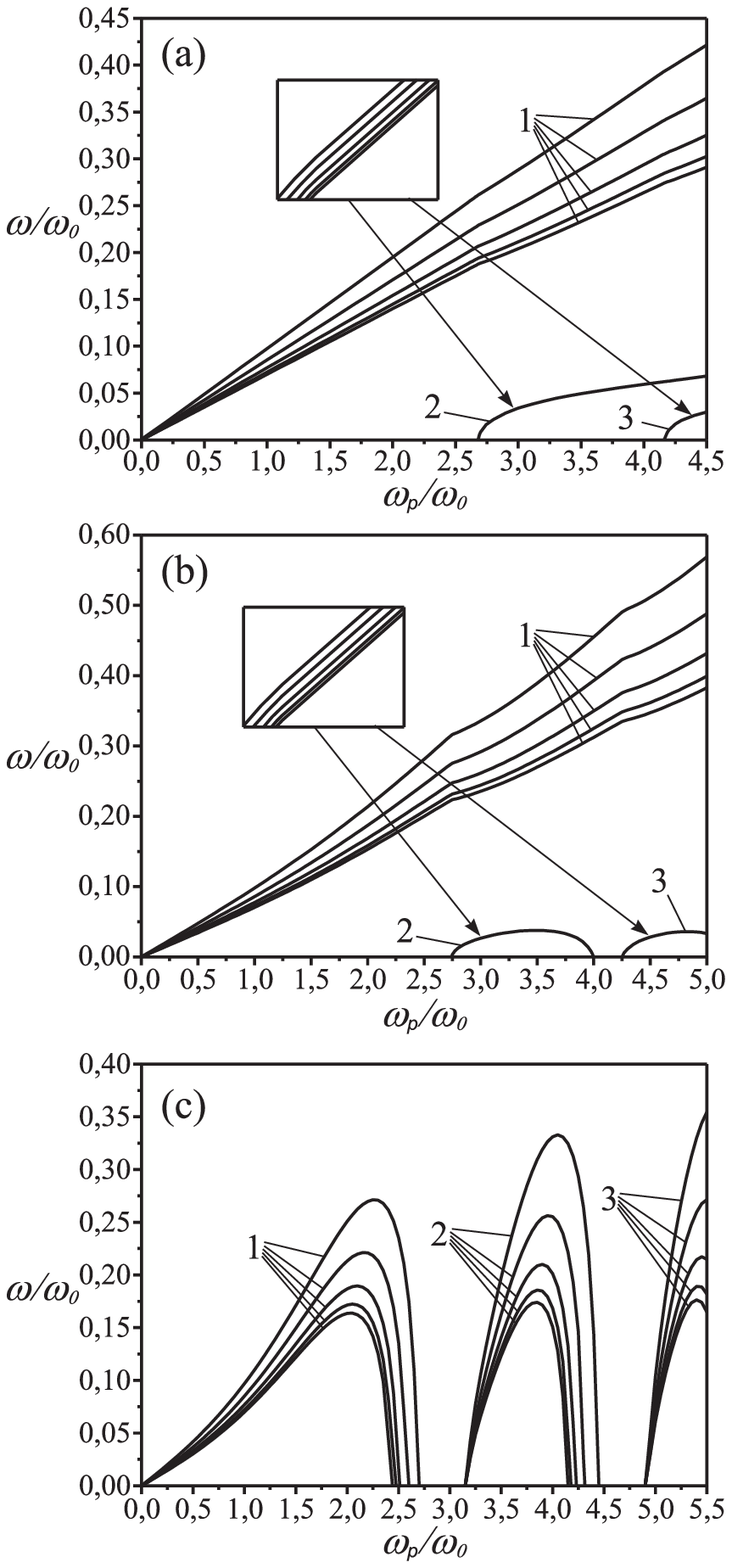}
   \end{tabular}
   \end{center}
\end{figure}
\begin{center} Figure 2. \end{center}

\pagebreak

\begin{figure}[t!]
   \begin{center}
   \begin{tabular}{c}
   \includegraphics[291,137][533,495]{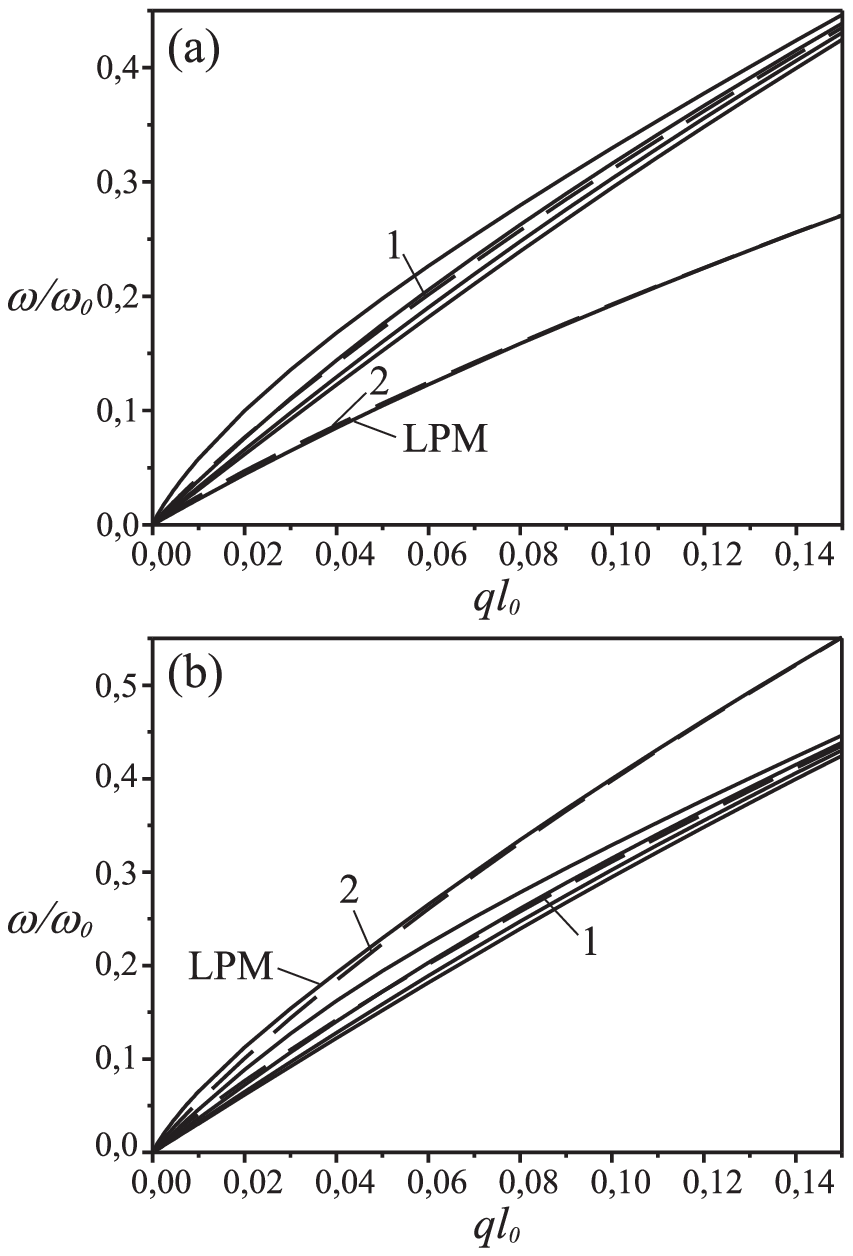}
   \end{tabular}
   \end{center}
\end{figure}
\begin{center} Figure 3. \end{center}

\pagebreak

\begin{figure}[t!]
   \begin{center}
   \begin{tabular}{c}
   \includegraphics[99,124][345,623]{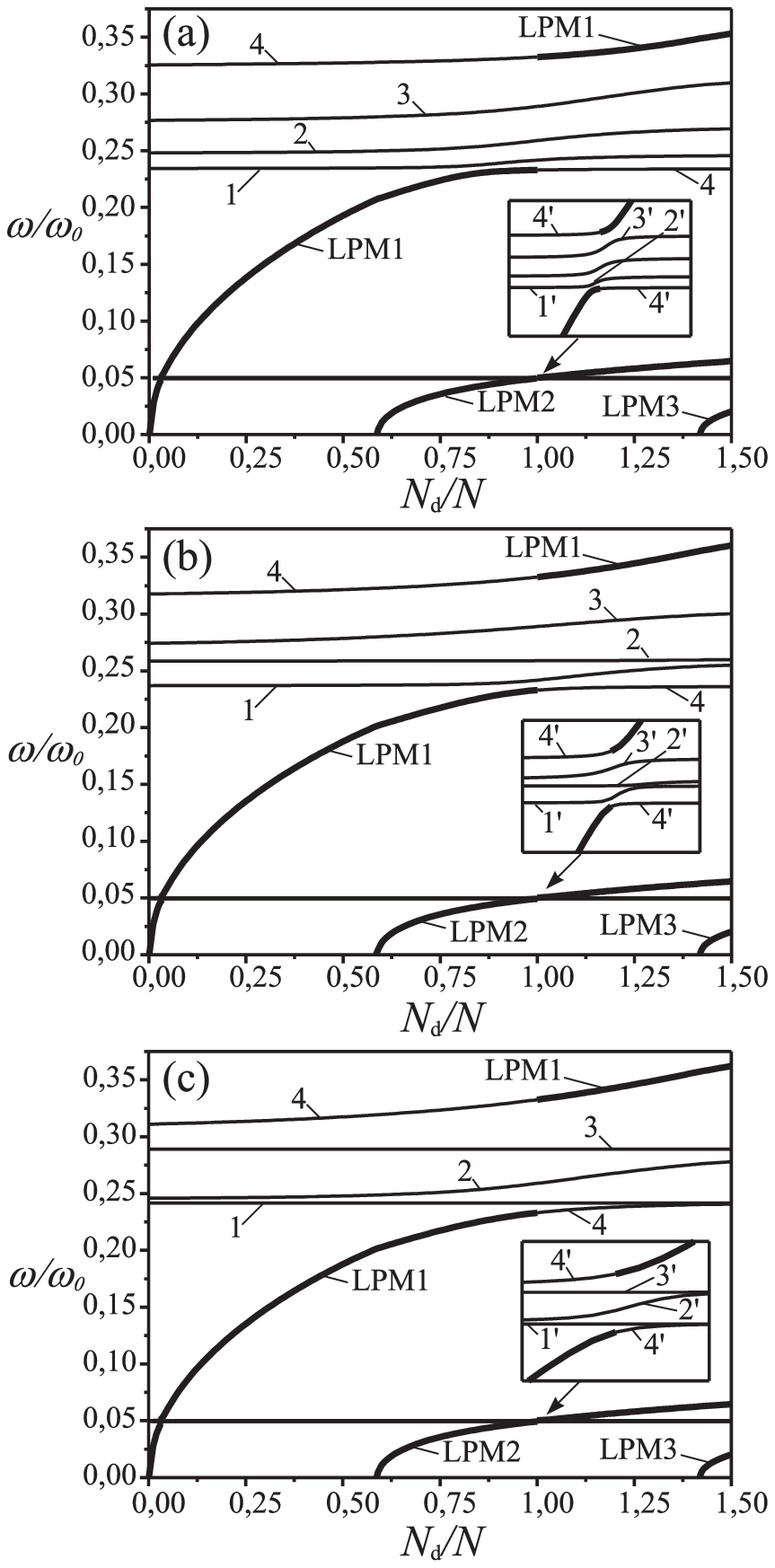}
   \end{tabular}
   \end{center}
\end{figure}
\begin{center} Figure 4. \end{center}

\pagebreak

\begin{figure}[t!]
   \begin{center}
   \begin{tabular}{c}
   \includegraphics[177,152][421,674]{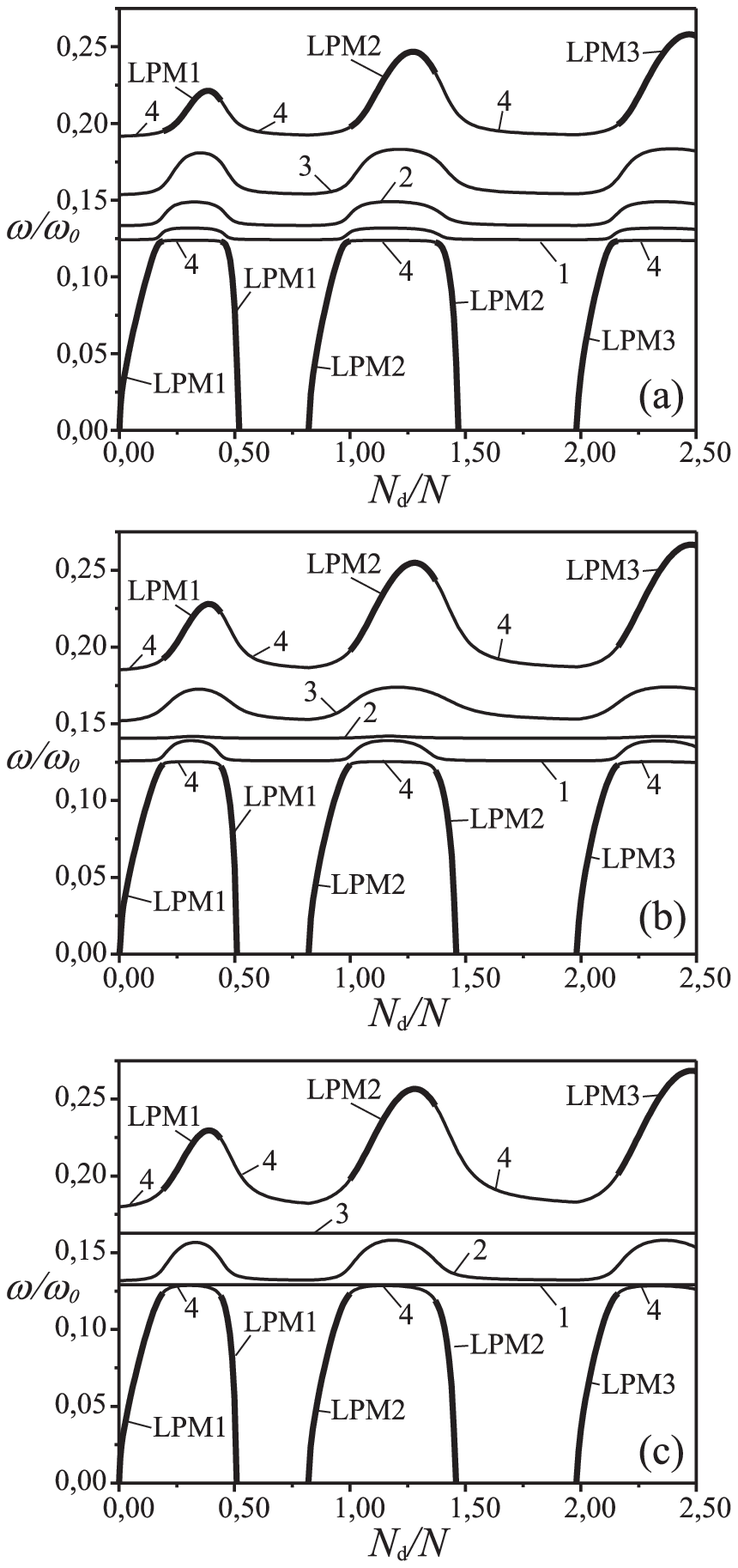}
   \end{tabular}
   \end{center}
\end{figure}
\begin{center} Figure 5. \end{center}

\end{document}